\documentclass[journal]{IEEEtran}
\usepackage[dvips]{graphicx}

\begin{document}
\title{Scintillation detectors constructed with an optimized 2x2 silicon
photomultiplier array}
\author{Felix~Liang,~\IEEEmembership{Member,~IEEE,}
  Hartmut~Brands,~\IEEEmembership{Member,~IEEE,}
  Les~Hoy,~\IEEEmembership{Member,~IEEE,}
  Jeff~Preston,~\IEEEmembership{Member,~IEEE,}
  and~Jason~Smith,~\IEEEmembership{Member,~IEEE,}
\thanks{Felix Liang, Hartmut Brands, Les Hoy, and Jason Smith are with FLIR
Systems Inc.}
\thanks{Jeff Preston, formerly with FLIR Systems Inc., is now with Consolidated
Nuclear Security, LLC.}
}
\maketitle

\begin{abstract}
Silicon photomultipliers (SiPMs) are a good alternative to photomultiplier
tubes (PMTs) because their gain and quantum efficiency are comparable
to PMTs. However, the largest single-chip
SiPM is still less than 1~cm$^2$. In order to use SiPMs with scintillators
that have reasonable sensitivity, it is necessary to use multiple
SiPMs. In this work, scintillation detectors are constructed and tested with
a custom 2x2 SiPM
array. The layout of the SiPMs and the geometry of the scintillator were
determined by performing Geant4 simulations. Cubic NaI, CsI, and CLYC with 18~mm
sides have been tested. The output of the scintillation detectors are stabilized
over the temperature range between --20 and 50~$^{\circ}$C by matching the
gain of the SiPMs in the array.
The energy resolution for these detectors has been measured
as a function of temperature.
Furthermore, neutron detection for the CLYC
detector was studied in the same temperature range. Using pulse-shape
discrimination, neutrons can be cleanly
identified without contribution from $\gamma$-photons.
As a result, these detectors are suitable for
deploying in spectroscopic personal radiation detectors (SPRD).
\end{abstract}


\section{Introduction}

Silicon photomultipliers (SiPMs) are constructed on a single substrate and
consist of thousands of microcells operating in the Geiger mode. The gain and
quantum efficiency of the SiPMs are comparable to those of the
photomultiplier tubes (PMTs). They are compact in size and
insensitive to magnetic fields. With low-operating voltages, it simplifies
the circuit design for electrical safety considerations. Therefore, they are
a good alternative to PMTs \cite{ec10}. However, the conventional PMTs are
available in various sizes that are as large as a few
tens of cm in diameter. In contrast, the largest single-chip SiPM is
less than 1~cm$^2$. For scintillation detectors, the efficiency of
scintillation photon collection increases with the area of the photon
sensor. Moreover, a larger detector volume has a higher sensitivity for
radiation detection. Because of the small active area of the SiPMs, it is
necessary
to use multiple SiPMs to increase the efficiency of photon collection for a
larger scintillator. In this work, scintillation detectors are
constructed and tested with a custom 2x2 SiPM array.

\section{Geant4 Simulations}
The layout of the SiPM array and the geometry of the scintillator were
determined by performing Geant4 simulations \cite{Geant4} in which a point
source is located 20~cm from the front of the scintillator. The
$\gamma$-rays are emitted uniformly random in a cone irradiating a circular
area inscribed by the front surface. The surface finish of the scintillator is
modeled using the ``groundteflonair'' option in the
Look-Up-Table (LUT) \cite{janecek}. A thin layer of
optical grease, 0.1~mm, is sandwiched between the scintillator and
the SiPMs for transporting scintillation photons.
The array has four 6x6 mm$^{2}$ SiPMs arranged in a 2x2
configuration such that the electronics is less complicated. The goal
of the simulations was to find the optimum geometry of the scintillator and
the layout of the SiPMs for good $\gamma$-ray energy resolution.

\begin{figure*}[!t]
\centering
\includegraphics[width=6in]{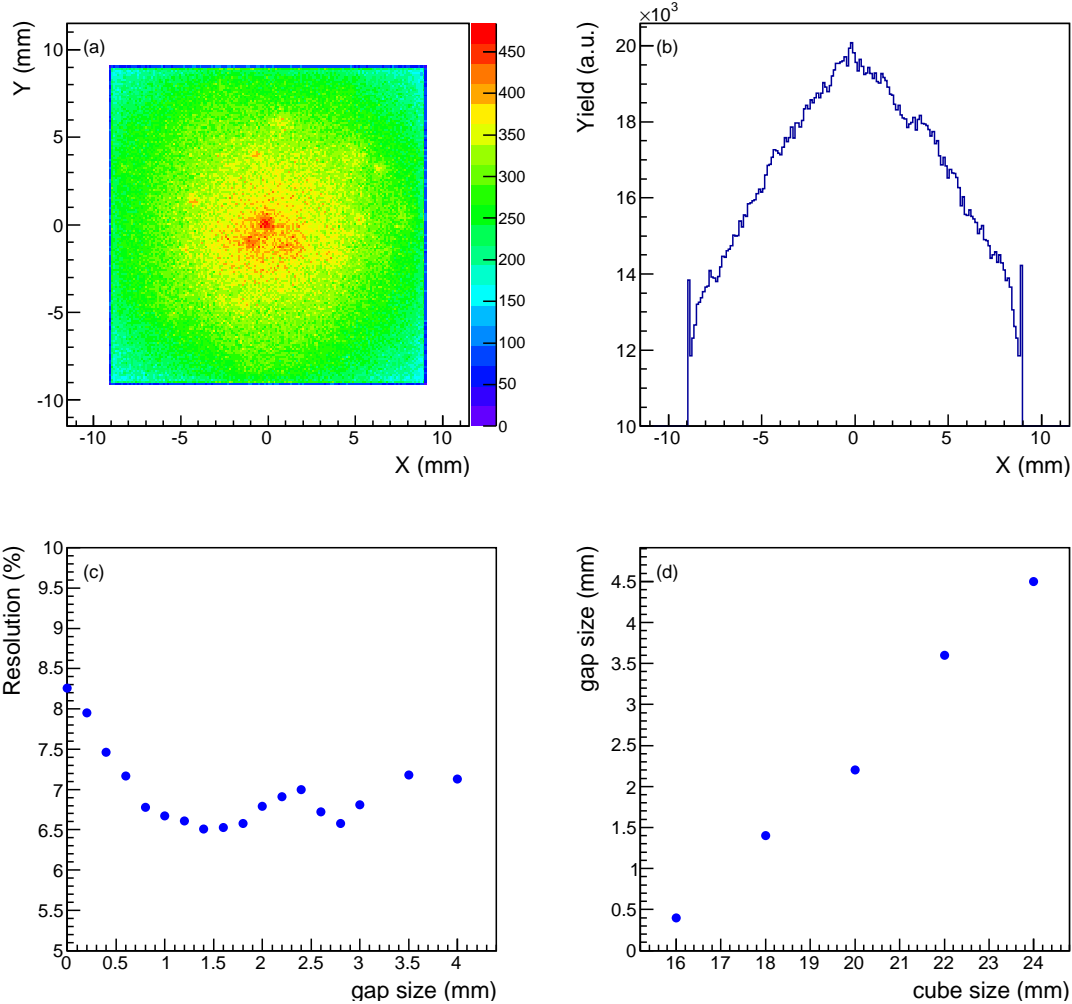}
\caption{Results of Geant4 simulations. (a) Distribution of scintillation
  photons at the exit of an 18~mm cubic scintillator irradiated by a 622~keV
  $\gamma$-ray located 20~cm away. (b) The profile of scintillation photon
  distributed in a 6~mm horizontal strip across the middle of the exit surface
  of the scintillator. (c) The detector resolution as a
  function of the gap size between the active area of SiPMs for a cubic NaI
  scintillator with 18~mm sides. (d) The gap size between the active area
  of SiPMs for scintillator cubes achieving an optimum energy resolution.}

\label{fg:resolgap}
\end{figure*}

It is found that the energy resolution of a detector is better for the
scintillator with an area comparable to the active area of the SiPM array.
Since the SiPMs are arranged as a square or rectangle in the array,
scintillators with a square or rectangular cross section have a better
energy resolution than those with a circular cross section. Shown in
Fig.~\ref{fg:resolgap}(a) is the photon distribution at the exit surface of
an 18~mm cubic scintillator. The profile of photon distribution for a 6~mm
horizontal strip across the middle of the exit surface is shown in
Fig.~\ref{fg:resolgap}(b). As can be seen, more scintillation photons are
distributed near the center of the exit surface of the scintillator.

\begin{figure}[!t]
\centering
\includegraphics[width=3.25in]{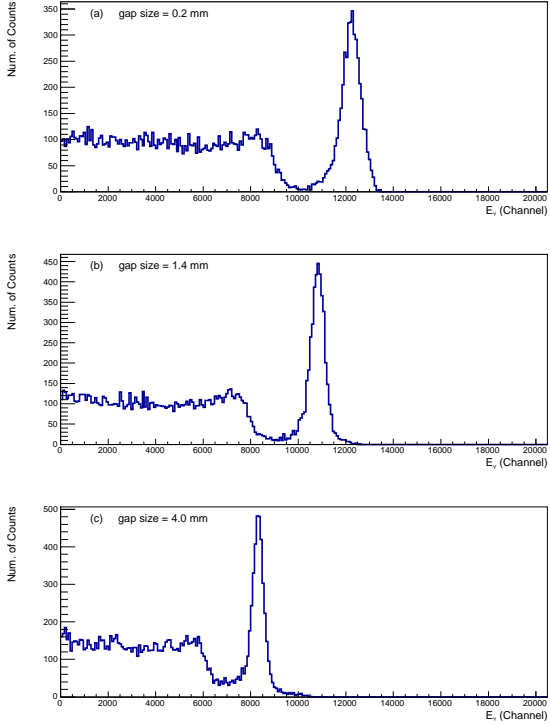}
\caption{Simulated spectra of a 662~keV $\gamma$-ray detected by an 18~mm cubic
  scintillator coupled to a 2x2 SiPM array. The gap size between the SiPMs is
  (a) 0.2, (b) 1.4, and (c) 4.0~mm.}
\label{fg:spk18gap}
\end{figure}

For scintillators with a cross section larger
than the active area of the SiPMs, leaving a gap between the SiPMs leads to a
better energy resolution.
In Fig.~\ref{fg:resolgap}(c), the energy resolution for an 18~mm cubic
scintillator as a function of the gap size between the active area of the
SiPMs is shown. The best energy resolution is for the gap size between 1 and
2~mm. For larger gap sizes,
the resolution gets worse slowly but a shoulder
starts to develop on either side of the photopeak resulting in a poorer
full-width-at-tenth-maximum. 
Fig.~\ref{fg:spk18gap} shows the simulated 662~keV $\gamma$-ray spectra
detected by the 18~mm cubic scintillator and SiPM array with
the gap size of 0.2, 1.4, and 4.0~mm.
It can be seen that for the gap size of 4.0~mm a noticeable shoulder appears
on the high-energy side of the photopeak and the trough between the photopeak
and the Compton edge is higher as well. Furthermore, the photopeak
position shifts to lower channels as the gap size increases. This is due to
less photons distributed away from the center of the detection plane
as shown in the profile of photon distribution in Fig.~\ref{fg:resolgap}(b). In
Fig.~\ref{fg:resolgap}(d), the gap size between the SiPMs for
cubic scintillators with dimensions between 16 and 24~mm to have an optimum
resolution is shown. As can be seen, it is necessary to increase the gap
size for larger scintillators in order to achieve a better energy resolution.

\section{Test of Scintillation Detectors}
A SiPM array with four 6x6~mm$^{2}$ SiPMs \cite{sensl} has been constructed
following the analysis of the Geant4 simulations. 
Three types of scintillators, NaI, CsI, and Cs$_2$LiYCl$_6$:Ce$^{3+}$(CLYC),
have been tested with this SiPM array. The
scintillator and SiPM array were enclosed in a hermetically sealed aluminum 
container to keep out moisture and ambient light. All the tests were
performed in a temperature controlled chamber.

\subsection{Detector Response}
The three detectors were tested using a $^{137}$Cs source with the SiPMs biased
at 27.5~V so that the detectors had the same nominal gain. The output of the
preamp was recorded by a waveform digitizer
(Struck SIS3302) operating at 100~MHz. The response of the detectors to the
662~keV $\gamma$-ray from $^{137}$Cs was
compared by making a histogram of the integral of the digitized pulses.
Because the CLYC pulse has a long-decay component, an integration time of
20~$\mu$s was used for all the scintillators in order to make an unbiased
comparison. Consequently, the counting rate was kept
low to minimize pulse pileup. As the decay time for NaI and CsI is
shorter, integrating these pulses for such a long time could introduce
noise to the integrals. Since the comparison was for the detector response,
the influence of noise on energy resolution was
ignored for the current test. A separate measurement using the proper
integration time for each scintillator was carried out to compare the
energy resolution.

\begin{figure}[!t]
\centering
\includegraphics[width=3.25in]{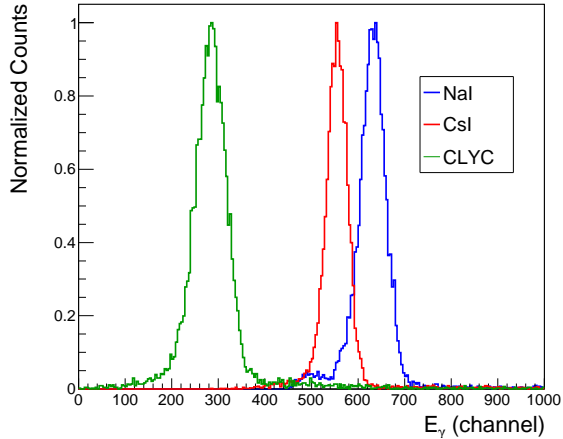}
\caption{The 662~keV photopeak of $^{137}$Cs detected by the NaI, CsI, and
  CLYC detectors. The histograms are obtained by integrating the
  pulses of the preamp output for 20~$\mu$s.}
\label{fg:lightout}
\end{figure}

Figure~\ref{fg:lightout} shows the normalized histogram of 10,000 pulse
integrals
for the three detectors. The NaI detector has the largest output because
the SiPM response is optimized for the 420~nm scintillation photon. Although
CsI has a larger yield of scintillation photons, the detector response is
actually
smaller than NaI. This is due to the mismatch between the scintillation
spectrum of CsI and the response of the SiPMs. Lastly, the light output for
CLYC is almost one half of that for NaI. Therefore, the pulse integral is the
smallest. 

\subsection{$\gamma$-ray Detection}
\begin{figure}[!t]
\centering
\includegraphics[width=3.25in]{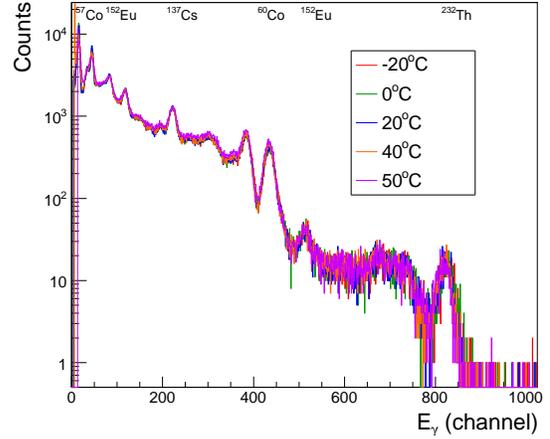}
\caption{Histograms of $\gamma$-rays from $^{152}$Eu,
  $^{60}$Co, $^{137}$Cs, and $^{232}$Th sources detected by the CsI detector
  at --20, 0, 20, and 50~$^{\circ}$C.}
\label{fg:det33msrc}
\end{figure}

The breakdown voltage for the SiPMs varies individually due to the
manufacturing processes. The breakdown voltage also increases with
temperature which results in gain variation following temperature changes.
In order to optimize the detector resolution, the gain for the
SiPMs was adjusted to be the same within the array. Furthermore, the gain for
the entire SiPM array was
stabilized over the temperature range between --20 and 50~$^{\circ}$C
\cite{pres}.
To check the quality of temperature stabilization, $\gamma$-ray sources,
$^{152}$Eu, $^{60}$Co, $^{137}$Cs, and $^{232}$Th, were used.
Fig.~\ref{fg:det33msrc} shows 
the spectra for these sources for the CsI detector
at --20, 0, 20, and 50~$^{\circ}$C. The centroid of the 662~keV photopeak
for $^{137}$Cs was set to channel 220 for all the temperatures.
As can be seen, the location of the photopeaks do not 
shift for different temperatures because of the stabilization performed for
the detector.

\begin{figure}[!t]
\centering
\includegraphics[width=3.25in]{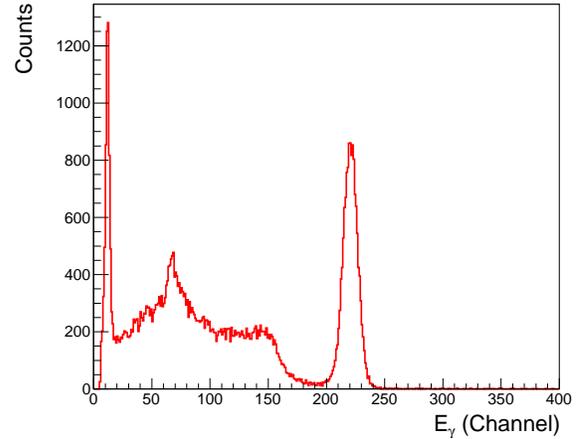}
\caption{Spectrum of $^{137}$Cs $\gamma$-rays detected by the CsI detector. The
gain of the SiPMs in the array has been matched to achieve an optimum
energy resolution of 6.4\%.}
\label{fg:cs137}
\end{figure}

To optimize the energy resolution, the preamp output was integrated
for 1, 4, and 20~$\mu$s for NaI, CsI, and CLYC, respectively.
The best resolution obtained was 6.8\% for NaI, 6.4\% for CsI, and 7.8\% for
CLYC. It has been reported that an energy resolution of 4\% was observed
for a 1~cm$^{3}$ CLYC scintillator coupled to a PMT.
However, when the same CLYC scintillator was coupled to
SiPMs the energy resolution was between 6.2\% and 8.3\% \cite{mes2016}.
The $^{137}$Cs $\gamma$-ray
spectrum measured by the CsI detector at 20~$^{\circ}$C is shown in
Fig.~\ref{fg:cs137}. The noise is sufficiently low such that the
32~keV peak is clearly visible without contamination.

\begin{figure}[!t]
\centering
\includegraphics[width=3.25in]{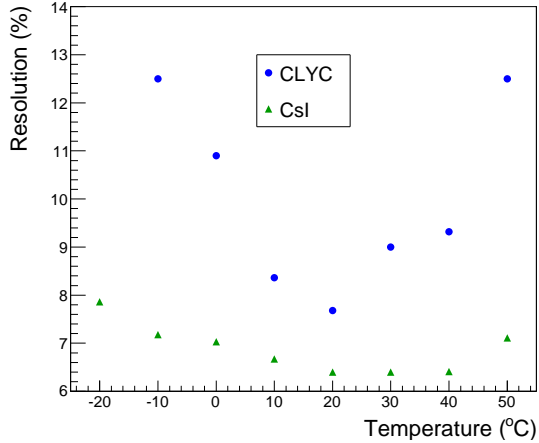}
\caption{(a) Energy resolution for CsI and CLYC detectors as a function of
  temperature.}
\label{fg:restemp}
\end{figure}

The energy resolution of the 662~keV photopeak for CsI and CLYC measured
between --20 and 50~$^{\circ}$C is displayed in Fig.~\ref{fg:restemp}.
The variation in energy resolution for the CsI detector is small. At
high temperatures, the poorer resolution is likely due to noise in the SiPMs
whereas at low temperatures the poorer resolution is attributed to a lower
light yield of the scintillator. For the CLYC detector, the energy resolution
is much worse than that for CsI, particularly well below and above room
temperature.
At --20~$^{\circ}$C, the 662~keV photopeak is barely visible in the
$\gamma$-ray spectrum. A previous study using a smaller crystal, 1~cm$^3$,
obtained mixed results for resolution as a function of temperature
\cite{mes2016}. In that study, SiPMs from different manufacturers
were compared. For the SensL SiPM, the variation of energy resolution
with temperature is smaller as compared to the present
work. For the Hamamatsu SiPM, resolution as poor as 14\% was measured at
10~$^{\circ}$C.
Since CLYC crystals are known to be fragile, it is conceivable that
fractures might have developed in the crystal that was used in this work. 
This makes the use of CLYC for portable radiation detectors challenging
because these instruments are required to operate between --20 and
50~$^{\circ}$C.

\begin{figure}[!t]
\centering
\includegraphics[width=3.25in]{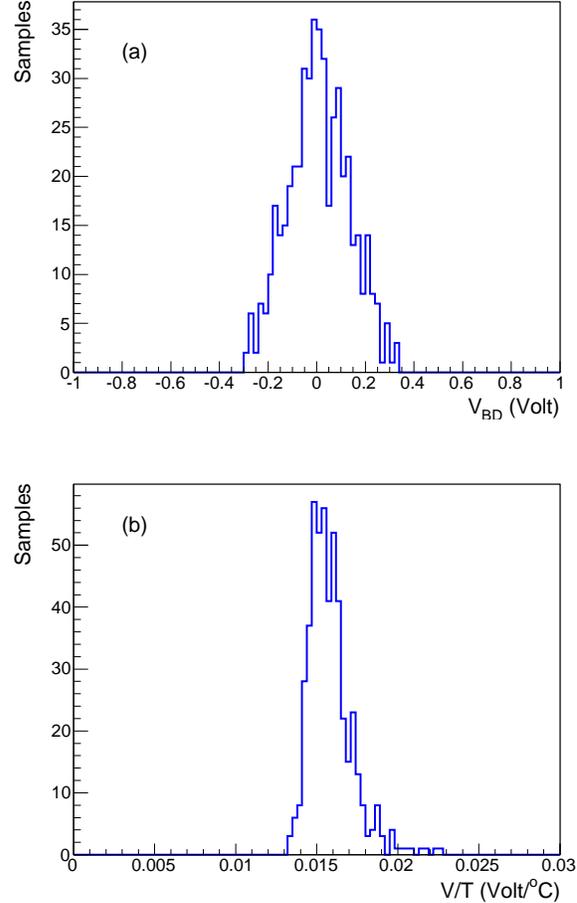}
\caption{(a) Distribution of the breakdown voltage for 300 SiPMs. (b)
Distribution of the temperature coefficient.}
\label{fg:distr}
\end{figure}

In addition to studying the three detectors,
a large number of CsI detectors have been tested which allow for examining
the characteristics of the SiPMs. Fig.~\ref{fg:distr}(a) shows the
distribution of the breakdown voltages for 300 SiPMs. The standard
deviation of the distribution is 0.124~V which is consistent with the
manufacturer's specification.
Fig.~\ref{fg:distr}(b) shows the distribution of the temperature coefficient
for the CsI detector. The centroid of the distribution is 15~mV/$^{\circ}$C.
In contrast, according to the manufacturer,
the temperature coefficient for the SiPM alone is 21.5~mV/$^{\circ}$C.

\subsection{Neutron Detection}
\begin{figure}[!t]
\centering
\includegraphics[width=3.25in]{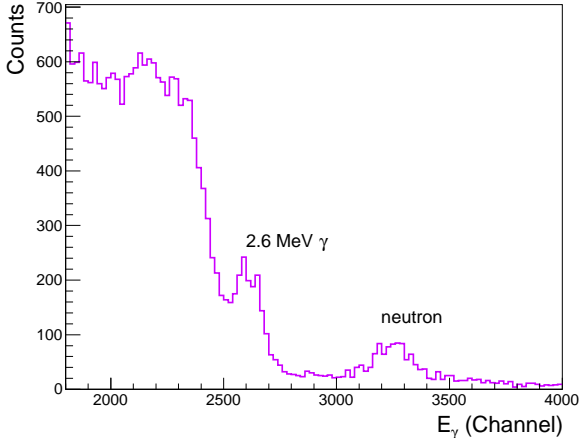}
\caption{Spectrum of $\gamma$-rays from $^{232}$Th and neutrons from $^{252}$Cf
detected by the CLYC detector.}
\label{fg:clyc1d}
\end{figure}

A $^{252}$Cf source was used for studying neutron detection by the CLYC
detector between --20  and 50~$^{\circ}$C. According to the specification, the
neutron is expected at 3.2~MeV in the $\gamma$-ray spectrum \cite{rmdclyc}.
Since $^{232}$Th has a 2.6~MeV $\gamma$-peak, it was used along with $^{252}$Cf
to test n-$\gamma$ discrimination and as a reference for the energy spectrum.
Because the largest difference in the pulse
shape occurs in the first 2~$\mu$s, a shorter integration time of 4.5~$\mu$s,
instead of 20~$\mu$s, was used to speed up the data
acquisition. Fig.~\ref{fg:clyc1d} shows the $\gamma$-ray spectrum detected by
the CLYC detector.
The photopeak for the $^{232}$Th 2.6~MeV $\gamma$-ray appears near
channel 2600 and the neutron peak appears near channel 3200. This agrees with
the manufacturer's specification \cite{rmdclyc}. As can be seen, the neutron
distribution is fairly broad and overlaps with the
high-energy tail of the 2.6~MeV $\gamma$-peak. For this reason,
it is necessary to use pulse-shape discrimination (PSD) to separate neutrons
from $\gamma$'s.

\begin{figure}[!t]
\centering
\includegraphics[width=3.25in]{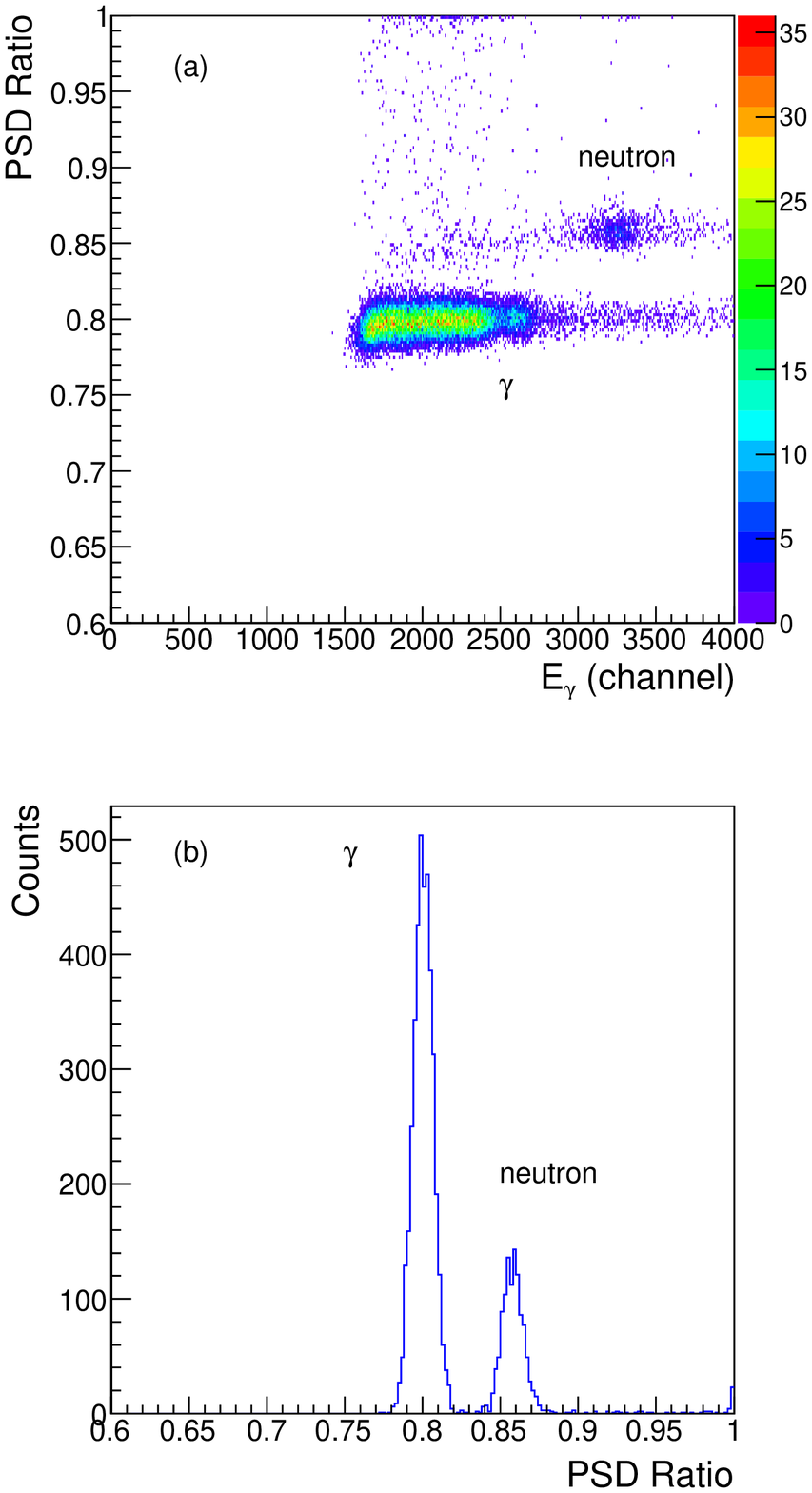}
\caption{(a) Histogram of PSD ratio versus pulse integral for the CLYC
detector. Two radiation sources, $^{252}$Cf and $^{232}$Th, were present during
the measurement. (b) Histogram of projected PSD ratio.}
\label{fg:cfthpsd20}
\end{figure}

The PSD is performed by taking the ratio of the
pulse integral in the delayed window to the sum of the pulse integral in
the prompt and delayed windows,
\begin{equation}
\centering
PSD\ Ratio = \frac{Delay}{Prompt+Delay}.
\end{equation}
Shown in Fig.~\ref{fg:cfthpsd20}(a) is the histogram for the PSD ratio versus
the integral of the preamp pulses. Neutrons appears in the group of counts in
the upper-right corner. The band of counts located across the center
of the histogram are $\gamma$-rays. The threshold for detection was set around
1.5~MeV to reduce the counting rate. The group of
events near the end of the $\gamma$ band on the right-hand side is the
2.6~MeV $\gamma$ peak from $^{232}$Th.

\begin{figure}[!t]
\centering
\includegraphics[width=3.25in]{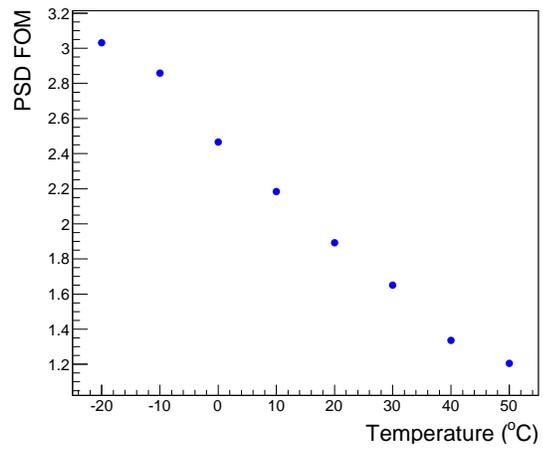}
\caption{The FOM of PSD for the CLYC detector as a function of temperature.}
\label{fg:fomtemp}
\end{figure}

By projecting the the
two-dimensional histogram on to the vertical axis, the histogram of the PSD
ratio is obtained, as shown in Fig.~\ref{fg:cfthpsd20}(b). The
figure-of-merit (FOM) for PSD is defined as the ratio of the difference
between the centroid ($\mu$) of the $\gamma$ and neutron peaks to the sum
of the full-width-at-half-maximum ($W$) of the two peaks \cite{mes2016},
\begin{equation}
\centering
FOM = \frac{\mu_{n}-\mu_{\gamma}}{W_{n}+W_{\gamma}}.
\end{equation}
The width for the prompt window and delayed window was varied to search for
the best FOM. At 20 $^{\circ}$C, the best FOM for n-$\gamma$ discrimination is
1.9 for the width of the prompt and delayed window of 270 and 850 ns,
respectively. The FOM increases with decreasing temperature due to the change
of pulse shape. 
At 50~$^{\circ}$C the FOM is 1.2 and at --20~$^{\circ}$C the FOM is 3.0, as
shown in Fig.~\ref{fg:fomtemp}. Although the resolution for CLYC is poor
at the extreme temperatures, the neutron identification by
pulse-shape discrimination works well.

\section{Conclusion}
The NaI and CsI scintillation detectors constructed with the custom SiPM array
have a good energy resolution for the 662~keV $\gamma$-ray from
$^{137}$Cs. They are suitable for deploying in a spectroscopic
personal radiation detector (SPRD) such as the FLIR identiFINDER R200.
Using pulse-shape discrimination, neutrons can be identified without
contamination from $\gamma$-rays for the CLYC detector.
However, the resolution for $\gamma$-ray detection
is poor at extreme temperatures, --20 and 50~$^{\circ}$C. Further work is
in progress to find solutions for neutron detection in hand-held radiation
detection instruments.

\end{document}